\documentclass[12pt]{article}
\usepackage{amsmath,amssymb}
\usepackage{graphicx}

\newcommand{\Z}{\mathcal{Z}}

\begin{document}

\thispagestyle{empty}
\begin{flushright}
OU-HET 674
\\
\end{flushright}
\vskip3cm
\begin{center}
{\Large {\bf Wall-crossing of D4-D2-D0 and flop of the conifold}}
\vskip3cm
{\large 
{\bf Takahiro Nishinaka\footnote{nishinaka [at] het.phys.sci.osaka-u.ac.jp}
 \,\,and\,\, Satoshi Yamaguchi\footnote{yamaguch [at] het.phys.sci.osaka-u.ac.jp}
}
}

\vskip1cm
{\it Department of Physics, Graduate School of Science, 
\\
Osaka University, Toyonaka, Osaka 560-0043, Japan}
\end{center}

\vskip1cm
\begin{abstract}
 We discuss the wall-crossing of the BPS bound states of a non-compact holomorphic D4-brane with D2 and D0-branes on the conifold. We use the Kontsevich-Soibelman wall-crossing formula and analyze the BPS degeneracy in various chambers.  In particular we obtain a relation between BPS degeneracies in two limiting attractor chambers related by a flop transition. Our result is consistent with known results and predicts BPS degeneracies in all chambers.
\end{abstract}


\newpage

\section{Introduction and conclusion}

Theories with extended supersymmetry have a special class of quantum states called BPS states. Their degeneracy (or index) is piecewise constant in the moduli space due to supersymmetry, but discretely changes when the moduli cross the ``walls of marginal stability.'' Since these walls are codimension one, the moduli space can be divided into chambers surrounded by marginal stability walls, and the degeneracy is exactly constant in each chamber.

Recently, there has been remarkable progress in the study of these wall-crossing phenomena of BPS states, especially in string theory on a Calabi-Yau three-fold. In the small string coupling regime, the BPS states are described by wrapped D-branes on supersymmetric cycles in the Calabi-Yau manifold. The wall-crossing of the BPS wrapped D-branes were studied from various points of view including the relation among statistical models, quiver gauge theory and the Donaldson-Thomas invariant \cite{Szendroi,NN,Nagao,Jafferis-Moore, Chuang-Jafferis, Ooguri-Yamazaki1, Dimofte-Gukov, Chuang-Pan, Sulkowski, DGS, Krefl, Aganagic-Schaeffer}, M-theory viewpoints \cite{AOVY, Aganagic-Yamazaki}, topological strings \cite{CSU, Nagao-Yamazaki}, and many others \cite{Diaconescu-Moore, Jafferis-Saulina, Andriyash-Moore, CDWM, Collinucci-Wyder, David, Manschot1, Herck-Wyder, Lee-Yi, Szabo:2009vw, CDP1, Manschot2, CDP2}.

The wall-crossing in string theory on a Calabi-Yau manifold was also studied from the four-dimensional ${\mathcal N}=2$ supergravity point of view. The appearance or disappearance of BPS states in the spectrum is related to the existence of the multi-centered BPS solutions in four-dimensional supergravity \cite{Denef:2000nb, Denef-Moore}. For example, for a two-centered BPS black hole, the distance between centers is changed in order to keep the BPS condition, if we move the moduli fields at spatial infinity. When the moduli at infinity cross the walls of marginal stability, the two-centered black hole ceases to exist due to the divergent distance. For some work on the wall-crossing phenomena from the supergravity viewpoint, see e.g., \cite{AdS3-S2,BEMB}.

Furthermore in \cite{Kontsevich-Soibelman} (see also \cite{Kontsevich-Soibelman2}), Kontsevich and Soibelman proposed a mathematical wall-crossing formula which tells us how the degeneracy of BPS states changes at the walls of marginal stability.\footnote{Primitive and semi-primitive wall-crossing formulae were already proposed in \cite{Denef-Moore} by supergravity analysis.} By using this formula, we can learn the BPS degeneracy in various chambers in the moduli space. For BPS states in ${\mathcal N}=2$ gauge theory, the physical meaning of this formula was studied in \cite{GMN1, GMN2, Cecotti-Vafa, GMN3}.

In this paper, we study the wall-crossing phenomena of D4-D2-D0 bound states on the resolved conifold by using the Kontsevich-Soibelman wall-crossing formula (KS formula), inspired by the work \cite{Jafferis-Moore} in which the wall-crossing of D6-D2-D0 bound states on the conifold was analyzed. The resolved conifold is a non-compact Calabi-Yau three-fold which has one compact two-cycle and no compact four-cycle. We introduce one D4-brane on a non-compact four-cycle of the conifold and evaluate the BPS degeneracy of the D2-D0 bound states on the D4-brane. By changing the K\"ahler moduli $z$ of the conifold, which are the size and the B-field for the compact two-cycle, various walls of marginal stability are crossed. 
Using the KS formula, we evaluate the partition functions in all the chambers in the moduli space.

In particular, there are two limits in the moduli space where all BPS states are realized in the field theory on D4-brane. We call them ``attractor chamber'' because they include attractor moduli of single-centered black holes. These two attractor chambers are related to each other by a topology-changing process called a ``flop'' transition of the conifold, which involves the changing of the intersection number of the four-cycle wrapped by the D4-brane and  the compact two-cycle in the conifold. 

\begin{figure}
\begin{center}
 \includegraphics[height=3.5cm]{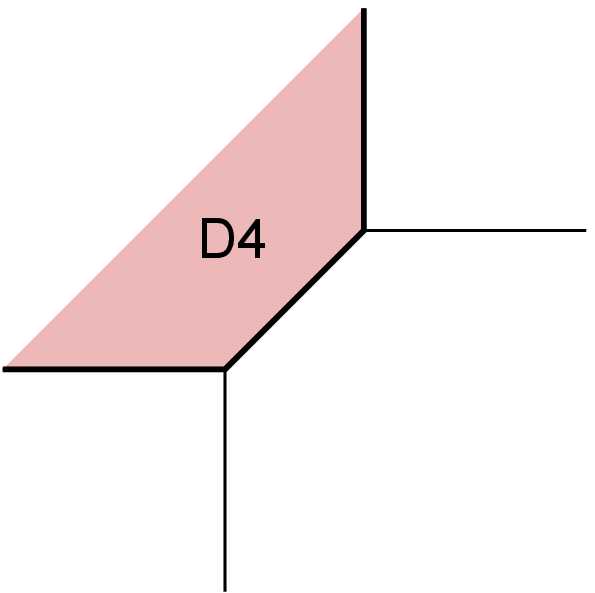}
 \hspace{2cm}
 \includegraphics[height=3.5cm]{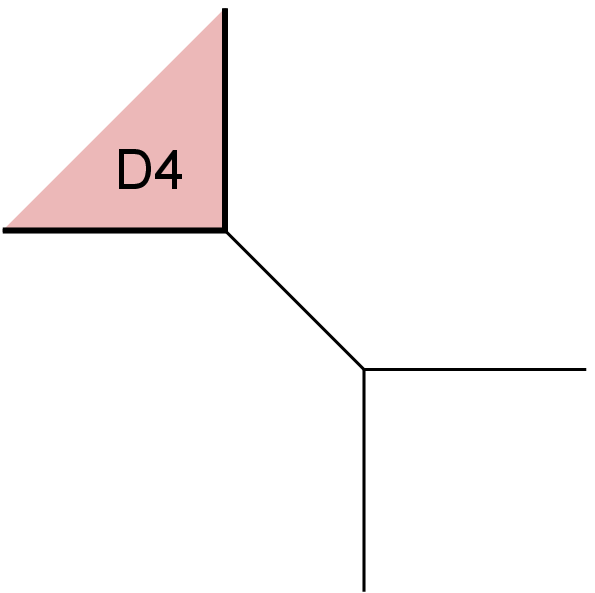}
\caption{Toric web diagrams of the conifolds and our D4-branes. The two figures are related by a flop transition.}
\label{fig:flop}
\end{center}
\end{figure}

In the language of toric web diagrams, the flop transition can be shown as in Fig.~\ref{fig:flop}.
Suppose that we move the K\"ahler parameter $z$ from ${\rm Im}\,z = +\infty$ to ${\rm Im}\,z = -\infty$. When ${\rm Im}\,z=0$, the size of the compact two-cycle becomes zero and the topology of the conifold changes. However, no singularity occurs in the theory here if we fix ${\rm Re}\,z\not\in \mathbb{Z}$. Two moduli regions of ${\rm Im}\,z >0$ and ${\rm Im}\,z<0$ correspond to the left and right pictures in Fig \ref{fig:flop}, respectively. The attractor chambers previously mentioned are ${\rm Im}\,z = \pm \infty$, which is shown by supergravity analysis. According to this observation, we study the wall-crossing of the partition function through the flop transition. By using the KS formula, we derive the relation between the partition functions in the two attractor chambers. This result is completely consistent with known facts \cite{AOSV, Vafa-Witten, Vafa1} about the partition function of the field theory on D4-branes.

Our result indicates that there is a structure of an infinite dimensional Lie algebra (affine SU(2) in this case, see eq.~\eqref{relation} for example) in the problem of BPS indices, which was also suggested in \cite{Cheng-Verlinde}. Actually the character of the affine Lie algebra appears in the instanton counting problem \cite{Nakajima,Vafa-Witten}. More detailed analysis in this direction is an interesting future problem.

Another interesting future problem will be similar analyses in other local Calabi-Yau manifolds.  Various kinds of topology changes of D4-branes will appear in these problems.
Making a statistical model, like crystals in the D6-D4-D2-D0 system, will be another interesting direction. The representation of the D4 partition function in terms of a sum over Young diagrams is well-known. Another example of such a statistical model is one in \cite{Jafferis:2006ny}.

This paper is organized as follows. In Section \ref{sec:wall} we consider the wall-crossing of D4-D2-D0 bound states on the resolved conifold and identify the walls of marginal stability. In Section \ref{sec:WCF} the wall-crossing formula for the partition function in our setup is derived by using the KS formula.
In Section \ref{section:attractor} the partition function in the attractor chamber is evaluated. We relate it to the partition function of the field theory on a D4-brane, by using the attractor mechanism in four-dimensional supergravity. The result is consistent with the wall-crossing formula derived in the previous section and the flop transition of the conifold.

\section{Local Calabi-Yau and walls of marginal stability}
\label{sec:wall}
In this section, we will set up the notation, and identify the locations of the walls of marginal stability.  We will follow the notation of \cite{Jafferis-Moore} and \cite{Denef-Moore} unless otherwise stated.

We consider the wall-crossing of D4-D2-D0 bound states on the resolved conifold ${\mathcal O}(-1)\oplus {\mathcal O}(-1) \to \mathbb{P}^1$ which is realized as the local limit of a compact Calabi-Yau three-fold. The BPS states we consider are composed of one non-compact D4-brane and an arbitrary number of D2-D0 states on it. We first put the D4-brane on a non-compact supersymmetric cycle ${\mathcal O}(-1)\to \mathbb{P}^1$ in the conifold. However if we consider a flop transition of the conifold, the topology of the four-cycle is changed. After the flop, the D4 brane is wrapped on the whole fiber directions and localized on the rigid $\mathbb{P}^1$. We will discuss this transition in section \ref{section:attractor}.

The charges of these BPS states are expressed in terms of even-forms as $\gamma = \mathcal{D}  +l_2\beta - l_0 dV$, where ${\mathcal D} \in H^{2}(X),\, {\mathcal \beta}\in H^{4}(X)$ and $dV \in H^6(X)$.  ${\mathcal D}$ represents the D4-brane charge while $l_2$ and $l_0$ denote the D2 and D0 charges respectively. We use the normalization of
\begin{eqnarray}
\int \mathcal{P}\wedge \beta = 1,
\quad \int \mathcal{P}'\wedge \beta =0,
\quad \int dV = 1
\end{eqnarray}
where $H^{2}(X)\ni \mathcal{P}$ is the basis of the compact harmonic 2-form, $H^{2}(X)\ni \mathcal{P}'$ is dual to the large four-cycle and $H^4(X)\ni \beta$ is dual to the compact 2-cycle. We can set $\mathcal{D}$ to satisfy $\mathcal{D}\cdot \mathcal{P}'^2 = 2c>0$ without loss of generality.  One also finds $\mathcal{D}\cdot \beta=-1$ because the D4-brane is wrapped on the 4-cycle $\mathcal{O}(-1)\to \mathbb{P}^1$ (see Figure \ref{fig:flop}).

We can write the complexified K\"ahler parameter of the conifold as $t = z{\mathcal P} + \Lambda e^{i\varphi}{\mathcal P}'$ where $z$ denotes the K\"ahler parameter for the rigid $\mathbb{P}^1$. The second term denotes the K\"ahler parameter for other non-compact cycles, and the local limit $\Lambda \to + \infty$ should be taken in the final result \cite{Jafferis-Moore}. So the moduli space we are interested in is the complex one-dimensional space of $z$. In the moduli region ${\rm Im}\,z > 0$, our D4-brane is wrapped on ${\mathcal O}(-1)\to \mathbb{P}^1$, while in the region ${\rm Im}\, z<0$ it stretches along the fiber directions and is localized on $\mathbb{P}^1$. Since our Calabi-Yau manifold is non-compact, the central charge of the BPS state with charge $\gamma = \mathcal{D} + l_2\beta - l_0 dV$ is evaluated as
\begin{align}
 Z(\gamma) = -\int_X \gamma\wedge e^{-t}  \sim -c\Lambda^2 e^{2i\varphi}
\end{align}
up to a real positive prefactor.\footnote{This prefactor is now irrelevant because, as is shown below, we only need to evaluate the phase of the central charge in order to identify the walls of marginal stability.}

Now let us turn to the walls of marginal stability and their locations.
The marginal stability walls are defined as a codimension one subspace in the moduli space where the BPS states can decay into other BPS states. Such a decay can occur only if the phases of the central charges of all the BPS states involved in the decay are the same.
Suppose that the state with charge $\gamma$ splits into states with $\gamma_2 = a + \mathcal{D}_h +\beta_h - n_h dV$ and $\gamma_1 = \gamma - \gamma_2$. The central charge of the state with $\gamma_2$ is
\begin{eqnarray}
 Z(\gamma_2)
 &\sim& \frac{a}{6}\Lambda e^{3i\varphi} - \frac{1}{2}\int \left(z\mathcal{P} + \Lambda e^{i\varphi}\right)^2\wedge \mathcal{D}_h - M_h\Lambda e^{i\varphi} + m_h z + n_h,
\end{eqnarray}
where $M_h = \beta_h\cdot \mathcal{P}'$ and $m_h=\beta_h \cdot \mathcal{P}$.
We can identify the walls of marginal stability of this D4-D2-D0 system as a subspace in the moduli space where $\arg Z(\gamma_2) = \arg Z(\gamma)$ is satisfied. By physical observation, we can assume $a=0$ and $\mathcal{D}_h=0$ or $\mathcal{D}$. We set $\mathcal{D}_h=0$ without loss of generality.
Furthermore, when $M_h\neq 0$, we have $Z(\gamma_2) \sim -M_h \Lambda e^{i\varphi}$ and $Z(\gamma) \sim -c\Lambda^2 e^{2i\varphi}$, which are never aligned in the non-compact limit. So we do not need to consider these walls.

Therefore, we only consider the case
\begin{eqnarray}
 \gamma = \mathcal{D} +l_2\beta - l_0dV \quad \longrightarrow \quad \left(\gamma_2 = m\beta -n dV\right) \,\,+\,\, \left(\gamma_1 = \gamma-\gamma_1\right).
\end{eqnarray}
The walls of marginal stability for these decay channels can be read off by comparing phases of $Z(\gamma) \sim -1/2\Lambda^2e^{2i\varphi}$ and $Z(\gamma_2) = mz +n$. Namely, the locations of the walls are given by
\begin{eqnarray}
 \varphi &=& \frac{1}{2}{\rm arg}\left(-mz-n\right) + \pi k,\qquad k\in \mathbb{Z}.\label{position_of_walls}
\end{eqnarray}
Note here that this condition is independent of the total D2/D0 charges $l_2$ and $l_0$.
Furthermore, one finds that we only have to consider the walls of $m=0,\pm 1$ because the only non-vanishing degeneracies of D2-D0 bound states on the resolved conifold are known to be \cite{Gopakumar-Vafa1, Gopakumar-Vafa2}
\begin{eqnarray}
\Omega(\pm \beta + ndV) =1,\quad \Omega(ndV) = -2. \label{D2-D0}
\end{eqnarray}
We can therefore classify the relevant walls of marginal stability into
\begin{eqnarray}
W_n^{m=-1}  &=& \{(z,\varphi) \,:\, \varphi = \frac{1}{2}{\rm arg}(z-n) + \pi k\},
\label{wall1} \\[1mm]
W_n^{m=+1}  &=& \{(z,\varphi) \,:\, \varphi = \frac{1}{2}{\rm arg}(-z-n) + \pi k\},
\label{wall2}\\[2mm]
W_{n}^{m=0}  &=& \{(z,\frac{\pi}{2})\}. 
\label{wall3}
\end{eqnarray}

In the next section, we will consider how the BPS indices change when the moduli cross these walls.

\section{Wall-crossing formula}
\label{sec:WCF}
\begin{figure}
 \begin{center}
  \includegraphics[width=10truecm]{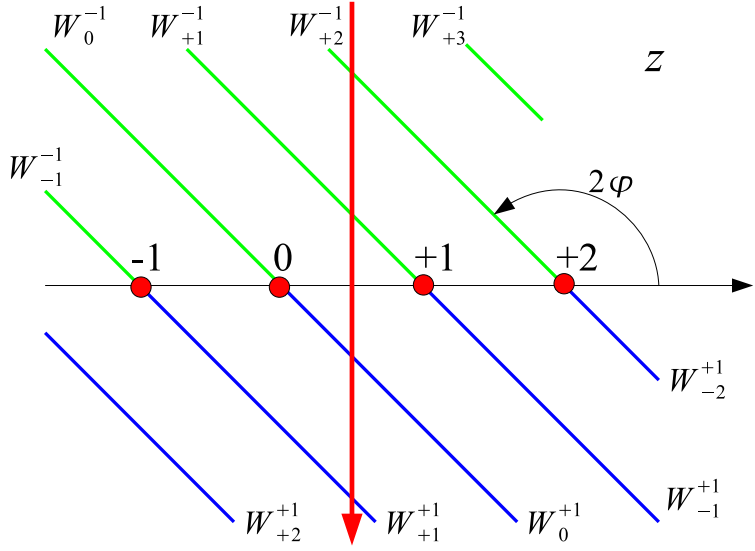}
 \end{center}
\caption{Walls on the $z$-plane with fixed $\varphi$. On the upper half plane, the walls $W^{-1}_{n}$ (green) are relevant, while the walls $W^{+1}_{n}$ (blue) are relevant on the lower half plane. We consider here the path $z=1/2+ia,\ a\in \mathbb{R}$ (red vertical line at the center).}
\label{fig:wall}
\end{figure}

We here briefly review the so-called Kontsevich-Soibelman wall-crossing formula (KS formula) \cite{Kontsevich-Soibelman} and study the change in the partition function of the D4-D2-D0 bound states under the wall-crossing.

The KS formula tells us that when the Calabi-Yau moduli $t$ cross the walls of marginal stability, the degeneracy $\Omega(\gamma;t)$ changes but the product
\begin{eqnarray}
 A &=& \prod^{\longrightarrow}_{\gamma=n\gamma_1+m\gamma_2,\,m>0,n>0} U_\gamma^{\Omega(\gamma;t)} \label{KS-product}
\end{eqnarray}
in the decreasing order of $\arg Z(\gamma)$ is unchanged. Here $U_\gamma = \exp\sum_{n=1}^\infty \frac{1}{n^2} e_{n\gamma}$ is defined in terms of generators $e_{\gamma}$ of an infinite-dimensional Lie algebra with the commutation relation
\begin{eqnarray}
\left[e_{\gamma_1},\, e_{\gamma_2}\right] &=& \left(-1\right)^{\left<\gamma_1,\gamma_2\right>}\left<\gamma_1, \gamma_2\right> e_{\gamma_1+\gamma_2}.
\end{eqnarray}
The ordering of the product in eq. \eqref{KS-product} does depend on the moduli $t$. The degeneracy $\Omega(\gamma,t)$ also depends on the moduli $t$. However, $A$ is independent of the moduli. Thus, the moduli dependence of $\Omega(\gamma;t)$ can be read off from the invariance of $A$ at the marginal stability walls.

When crossing the wall $W_n^{m}$ defined in eq.~\eqref{wall1}-\eqref{wall3} where the decay $\gamma \to \gamma_1 + \gamma_2$ can occur, the sign of $\arg Z(\gamma) - \arg Z(\gamma_2)$ is changed. Correspondingly, the order of $(\prod_\gamma U_{\gamma}^{\Omega(\gamma)})$ and $(\prod_{j=1}^\infty U_{j\gamma_2}^{\Omega(j\gamma_2)})$ is reversed in $A$.\footnote{To be more precise, all charge combinations of the form $k\gamma + j\gamma_2\,\,(k,j=0,1,2,\cdots)$ are involved in the wall-crossing at $W_n^m$. However, since we are only interested in the degeneracy of the BPS states that have only one D4 brane charge, it is sufficient to consider only the charges of $k=0$ and $k=1$. Furthermore, for the charge of $k=1$, we only need to consider the case $j=0$ when the product over $\gamma = {\mathcal D} + l_2\beta - l_0 dV$ is taken into account. Thus we finally obtain eq.~\eqref{wcf1}.} Recalling the invariance of the product $A$ under the wall-crossing, we find the following equation
\begin{align}
 \left(\prod_{\gamma}U_{\gamma}^{\Omega(\gamma)}\right)\left(\prod_{j=1}^{\infty}U_{j\gamma_2}^{\Omega(j\gamma_2)}\right)
= \left(\prod_{j=1}^{\infty}U_{j\gamma_2}^{\widetilde\Omega(j\gamma_2)}\right)\left(\prod_{\gamma}U_{\gamma}^{\widetilde\Omega(\gamma)}\right).
\label{wcf1}
\end{align}
Here $\Omega(\gamma)$ and $\widetilde\Omega(\gamma)$ denote the BPS degeneracy before wall-crossing and after wall-crossing, respectively. From this equation, we can read off the relation between $\Omega$ and $\widetilde{\Omega}$. First, by comparing the coefficients of $e_{j\gamma_2}$ in eq.~\eqref{wcf1}, one finds
\begin{align}
 \widetilde\Omega(j\gamma_2)=\Omega(j\gamma_2).
\label{gamma2}
\end{align}
Next let us expand $\left(\prod_{\gamma}U_{\gamma}^{\Omega(\gamma)}\right)$ and collect the terms $e_{\gamma}$ with $\gamma = \mathcal{D} +l_2\beta - l_0dV$. Then one obtains
\begin{align}
 \left(\prod_{\gamma}U_{\gamma}^{\Omega(\gamma)}\right)
=1+\sum_{\gamma}\Omega(\gamma)e_{\gamma}+\dots.
\end{align}
Thus eq.~\eqref{wcf1} with eq.~\eqref{gamma2} reads
\begin{align}
 \sum_{\gamma}\widetilde\Omega(\gamma)e_{\gamma}&=
\left(\prod_{j=1}^{\infty}U_{j\gamma_2}^{-\Omega(j\gamma_2)}\right)\sum_{\gamma}\Omega(\gamma)e_{\gamma}\left(\prod_{j=1}^{\infty}U_{j\gamma_2}^{\Omega(j\gamma_2)}\right)\nonumber\\
&=\sum_{\gamma}\Omega(\gamma)e_{\gamma}\circ \prod_{j=1}^{\infty}\circ (1+(-1)^{j\langle\gamma_2,\gamma\rangle}e_{j\gamma_2})^{\circ j\langle\gamma_2,\gamma\rangle \Omega(j\gamma_2)},\label{wcf2}
\end{align}
where $\circ$ denotes the commutative product defined as
\begin{align}
 e_{\gamma_1}\circ e_{\gamma_2}:=e_{\gamma_1+\gamma_2}.
\end{align}
Eq. \eqref{wcf2} gives the simple expression of the wall-crossing formula for the partition functions. Indeed, by defining the partition functions as
\begin{align}
 \Z(u,v)=\sum_{\ell_2,\ell_0}\Omega(\mathcal{D} +l_2\beta - l_0dV)u^{\ell_0}v^{\ell_2},\qquad
 \widetilde \Z(u,v)=\sum_{\ell_2,\ell_0}\widetilde\Omega(\mathcal{D} +l_2\beta - l_0dV)u^{\ell_0}v^{\ell_2},
\end{align}
the wall-crossing formula \eqref{wcf2} can be written as
\begin{align}
 \widetilde{\Z}(u,v) =& \Z(u,v)\times\prod_{j=1}^\infty\left(1+\left(-1\right)^{j\left<\gamma_2,\gamma\right>}(u^{n}v^{m})^j\right)^{j\left<\gamma_2,\gamma\right>\Omega(j\gamma_2)},\label{KS}
\end{align}
where $m,n$ denote the D2 and D0 charges of $\gamma_2$ i.e. $\gamma_2=m\beta-ndV$.\footnote{This is the semi-primitive wall-crossing formula first obtained in \cite{Denef-Moore}.}

In the following we apply the formula \eqref{KS} to our current problem. Let us fix $\varphi$ to a certain value in $\pi/4<\varphi<\pi/2$ and explore the $z$-plane (Fig.~\ref{fig:wall}). Note that the relevant walls are only \eqref{wall1} and \eqref{wall2}. The walls of \eqref{wall3} do not give rise to any jump in the partition function since we have no D6-brane.
We also fix the real part of $z$ to be $1/2$ for simplicity, and denote the imaginary part of $z$ by $a$.  In other words, we consider the path
\begin{align}
 z=1/2+ia,\qquad -\infty < a < +\infty,
\end{align}
as is indicated by the red vertical line in figure \ref{fig:wall}, and all the walls crossed by it.

Let $\Z_{a}(u,v)$ be the partition function for the chamber including $z=1/2+ia$.  When $a\geq 0$, the relevant walls are $W^{-1}_{n}\, (n=1,2,3,\dots)$. In the chamber between $W_{n-1}^{-1}$ and $W_{n}^{-1}$ for an arbitrary $n\geq 1$,  the partition function can be read off from the wall-crossing formula \eqref{KS} as
\begin{align}
 \Z_{a\geq 0}(u,v)=\Z_{+\infty}(u,v)\prod_{r=n}^{\infty}\frac{1}{(1-u^r v^{-1})},
\end{align}
where we used $\left<\beta,{\mathcal D}\right>=-1$ and the fact, which immediately follows from \eqref{D2-D0}, that $\Omega(j\gamma_2) = 0$ unless $j=1$.
In particular, when $a=0$ the partition function becomes
\begin{align}
 \Z_0(u,v)=\Z_{+\infty}(u,v)\prod_{r=1}^{\infty}\frac{1}{(1-u^r v^{-1})}.
\label{positive a}
\end{align}
On the other hand, we find the relevant walls are $W^{+1}_{n}\, (n=0,1,2,\dots)$ for $a\leq 0$. By a similar argument to the above, the partition function in the chamber between $W_{n-1}^{+1}$ and $W_{n}^{+1}$ for $n\geq 0$ is written as
\begin{align}
 \Z_{a\leq 0}(u,v)=\Z_{-\infty}(u,v)\prod_{r=n}^{\infty}(1-u^r v).
\end{align}
In particular, one finds that
\begin{align}
 \Z_{0}(u,v)=\Z_{-\infty}(u,v)(1-v)\prod_{r=1}^{\infty}(1-u^r v).
\label{negative a}
\end{align}
Of course, this should be the same partition function as \eqref{positive a}. From the equivalence of \eqref{positive a} and \eqref{negative a}, we can read off the relation between $\Z_{+\infty}$ and $\Z_{-\infty}$ as
\begin{align}
 \Z_{+\infty}(u,v)=\Z_{-\infty}(u,v)(1-v)\prod_{r=1}^{\infty}(1-u^r v)(1-u^r v^{-1}).\label{relation}
\end{align}

In summary we used the wall crossing formula and expressed the partition functions in all the chambers in terms of $\Z_{+\infty}$ and $\Z_{-\infty}$. In the next section we will consider the explicit form of $\Z_{\pm \infty}$.

\section{BPS degeneracy in the attractor chamber}
\label{section:attractor}

We here evaluate the explicit expressions of $Z_{\pm\infty}$. We first calculate $Z_{+\infty}$ and then read off $Z_{-\infty}$ from relation \eqref{relation}. The result is completely consistent with the flop transition of the conifold.

We first evaluate $\Z_{+\infty}$ by relating it to the partition function of the field theory on the D4-brane wrapped on $\mathcal{O}(-1) \to \mathbb{P}^1$.  In the previous section, we defined $\Z_{+\infty}$ as the partition function of  D4-D2-D0 bound states in the large $\mathbb{P}^1$ limit. So we now show that the partition function for the chamber including ${\rm Im}\,z = +\infty$ can be evaluated in the field theory on the D4-brane.

For this purpose, we use four-dimensional supergravity analysis. In the supergravity point of view, the wall-crossing of BPS states is interpreted as the disappearance or appearance of {\em multi-centered} black holes in the spectrum \cite{Denef:2000nb,Denef-Moore}. In the moduli space, there is a special chamber including ``attractor points.'' The attractor point is defined as the value of the moduli at a {\em single-centered} black hole horizon, which is completely determined by the electric and magnetic charges of the black hole \cite{FKS, Strominger, Ferrara:1996dd, Ferrara-Kallosh, FGK}. Therefore, when the moduli is in the chamber including the attractor points, which we call the ``attractor chamber,'' we expect that all the BPS states are realized as single-centered black holes. Furthermore, in our set up the microstates of the single-centered black holes can be counted in the field theory on D4-brane \cite{MSW}. So
for our purpose, it suffices to show that $\Z_{+\infty}$ is the partition function for the attractor chamber.

\begin{figure}
\begin{center}
\includegraphics[width=10truecm]{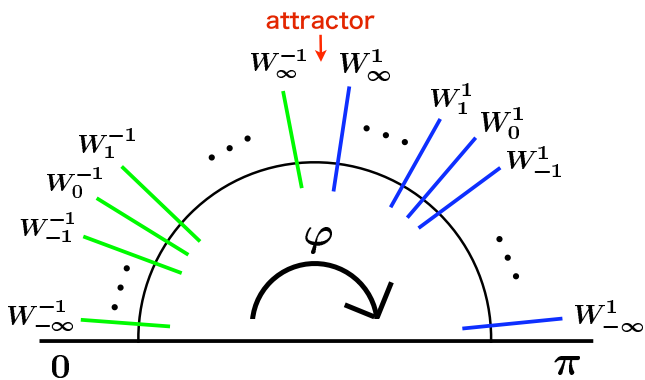}
\caption{The positions of the walls of marginal stability in the $\varphi$-plane with fixed $z$.}
\label{fig:chamber}
\end{center}
\end{figure}

The attractor point is determined by the charge through the attractor equation (see eq. (3.12) in ref. \cite{Denef-Moore})
\begin{eqnarray}
 2{\rm Im}(\overline{Z}(\gamma)\,{\bf \Omega}) = -\gamma,
\end{eqnarray}
where ${\bf \Omega}$ denotes the normalized period vector. By solving this for the charge $\gamma = \mathcal{D} + l_2 \beta - l_0 dV$, we obtain $\varphi = \pi/2$. Thus the attractor chamber is the one which includes $\varphi = \pi/2$. Note that this condition is independent of $z$, the K\"ahler parameter for the compact cycle. So we first identify the attractor chamber in the $\varphi$-plane with fixed $z$, and then we translate it in the $z$-plane with fixed $\varphi$. 
%
Recall that the relevant walls of marginal stability are eqs.~\eqref{wall1} and \eqref{wall2}. The walls in eq.~\eqref{wall3} do not give rise to any jump in the partition function because there is no D6-brane. Thus the position of the walls in the $\varphi$-plane are depicted as in Fig.~\ref{fig:chamber}.
The attractor chamber is the one between the walls $W_{\infty}^{1}$ and $W_{\infty}^{-1}$. On the other hand, in the $z$-plane with fixed $\varphi$, the walls have been drawn as in Fig.~\ref{fig:wall}. The attractor chamber is now found to exist at $z=\pm i\infty$. This implies that $\Z_{+\infty}$ is equivalent to the partition function for the attractor chamber.\footnote{We see that ${\rm Im}\,z = -\infty$ is another attractor chamber. Thus $\Z_{-\infty}$ is also related to the partition function of field theory on a D4-brane. In fact, it is related to $\Z_{+\infty}$ by the flop transition of the resolved conifold as mentioned below.
}

From the above arguments, we see that $\Z_{+\infty}$ is equivalent to the partition function of the field theory on a D4-brane wrapped on $\mathcal{O}(-1)\to\mathbb{P}^1$. In fact, it was evaluated in \cite{AOSV} (see also \cite{Vafa-Witten, Vafa1}). In our notation it is written as
\begin{eqnarray}
\Z_{+\infty}(u,v) &=& f(u)(1-v)\prod^\infty_{r=1}\left(1-u^r\right)\left(1-u^rv\right)\left(1-u^rv^{-1}\right),
\end{eqnarray}
where $f(u)$ is related to the bound states of D0-branes on the D4-brane without flux, which cannot be fixed because our D4-brane is non-compact. The degeneracy of the bound states with non-vanishing D2-brane charge is unambiguously determined. Note that without D2-branes, the flux on the D4-brane can induce D2-brane charge through the Chern-Simons interaction, because $\mathcal{O}(-1)\to \mathbb{P}^1$ has a compact two-cycle on it. Also recall that when we consider the theory on a compact D4-brane, we find $f(u) = \prod_{r=1}^{\infty}(1-u^r)^{-\chi(C_4)}$ with the Euler characteristic of the four-cycle $\chi(C_4)$. 

Finally, we mention $\Z_{-\infty}$ defined in the previous section. By fixing ${\rm Re}\,z \neq 0$ and moving the K\"ahler parameter from ${\rm Im}\, z = +\infty$ to ${\rm Im}\, z = -\infty$, the flop transition occurs. After the flop, the four-cycle wrapped by the D4-brane now has {\em no compact two-cycle} in it through which the flux on the D4-brane can induce the D2-brane charge. Furthermore, we have seen that ${\rm Im}\,z = -\infty$ is also an attractor chamber. Thus we expect that $\Z_{-\infty}$ is equivalent to the partition function of a field theory on D4-brane with no D2-brane charge contribution. In fact, from the wall-crossing formula (\ref{relation}), we can show
\begin{align}
 \Z_{-\infty}(u,v) = \Z_{+\infty}(u,v)\times \prod_{r=0}^\infty\left(1-u^r v\right)^{-1} \times \prod_{r=1}^\infty\left(1-u^r v^{-1}\right)^{-1}
\,\,=\,\,\, f(u)\,\prod_{r=1}^{\infty}(1-u^r).
\end{align}
This is independent of the chemical potential for D2-branes.
By recalling that $f(u)=\prod_{r=1}^{\infty}(1-u^r)^{-\chi(C_4)}$ for a compact D4-brane case, this result is also consistent with the fact that the Euler characteristic $\chi(C_4)$ decreases by one through the flop transition.


\section*{Acknowledgments}

We would like to thank Takahiro Kubota for many illuminating discussions, important comments and suggestions.
 We are grateful to Wade Naylor for a careful reading of this manuscript and useful comments.
We also wish to thank Tohru Eguchi, Kazuo Hosomichi, Yutaka Hosotani, Kyosuke Hotta, Tsuyoshi Houri, Sungjay Lee, Kazunobu Maruyoshi, Koichi Nagasaki, Toshio Nakatsu, Yu Nakayama, Tetsuya Onogi, Takeshi Oota, Noburo Shiba, Yuji Sugawara, Hiroaki Tanida and Yukinori Yasui for many useful discussions.
S.Y. would like to thank the organizers and the participants of ``Focus Week on New Invariants and Wall Crossing'' May 18--22, 2009 at IPMU, Kashiwa, where he benefitted from interesting talks and stimulating discussions, and got interested in this topic.
T.N. was supported in part by JSPS Research Fellowship for Young Scientists. 
S.Y. was supported in part by KAKENHI 22740165.

\end{document}